\documentclass[3p,times,procedia]{elsarticle}
\usepackage{nupha_ecrc}

\volume{00}

\firstpage{1}

\journalname{Nuclear Physics A}

\runauth{}

\jid{nupha}

\jnltitlelogo{Nuclear Physics A}

\usepackage{amssymb}

\usepackage[figuresright]{rotating}
\usepackage{subcaption}
\usepackage{lineno}

\begin{document}

\begin{frontmatter}




\title{Charm-hadron production in $pp$ and AA collisions}


\author{Min He}

\address{Department of Applied Physics, Nanjing University of Science and Technology, Nanjing 210094, China}

\author{Ralf Rapp}

\address{Cyclotron Institute and Department of Physics \& Astronomy,
Texas A\&M University, College Station, TX 77843, USA}

\begin{abstract}
Recent measurements of various charm-hadron ratios in $pp$, $p$-Pb and Pb-Pb collisions at the LHC have
posed challenges to the theoretical understanding of heavy-quark hadronization. The $\Lambda_c/D^0$ ratio
in $pp$ and $p$-Pb collisions shows larger values than those found in $e^+e^-$ and $ep$ collisions and predicted
by Monte-Carlo event generators based on string fragmentation, at both low and intermediate transverse momenta ($p_T$).
In AA collisions, the $D_s/D$ ratio is significantly enhanced over its values in $pp$, while the $\Lambda_c/D^0$ data
indicates a further enhancement at intermediate $p_T$. Here, we report on our recent developments for a comprehensive
description of the charm hadrochemistry and transport in $pp$ and $AA$ collisions. For $pp$ collisions we find
that the discrepancy between the $\Lambda_c/D^0$ data and model predictions is much reduced by using a statistical
hadronization model augmented by a large set of ``missing" states in the charm-baryon spectrum, contributing to
the $\Lambda_c$ via decay feeddown. For $AA$ collisions, we develop a 4-momentum conserving resonance recombination
model for charm-baryon formation implemented via event-by-event simulations that account for space-momentum
correlations (SMCs) in transported charm- and thermal light-quark distributions. The SMCs, together with the augmented
charm-baryon states, are found to play an important role in describing the baryon-to-meson enhancement at intermediate
momenta. We emphasize the importance of satisfying the correct (relative) chemical equilibrium limit when computing the
charm hadrochemistry and its momentum dependence with coalescence models.
\end{abstract}


\end{frontmatter}



\section{Introduction}
Heavy quarks ({\it i.e.}, charm, $c$, and bottom, $b$), ever since their discovery, have served as a versatile tool to
test Quantum Chromodynamics (QCD), mainly facilitated by their masses which are large compared to the nonperturbative scale
of QCD: $m_{c,b}\gg \Lambda_{\rm QCD}$. This renders their pairwise production ($c\bar{c}$ or $b\bar{b}$)  perturbative,
which must be followed, however, by a nonperturbative hadronization process at large distances. In $pp$ collisions, the
conventional hadronization mechanism for heavy quarks is fragmentation modeled by empirical functions. On the other hand,
in relativistic heavy-ion ($AA$) collisions, the formation of a  hot and dense deconfined Quark-Gluon Plasma
(QGP) enables heavy quarks to hadronize through recombination with nearby light (anti-) quarks in the medium.
Furthermore, since heavy quarks participate in the full evolution history of the fireball, via diffusion in the QGP,
hadronization and further diffusion in the hadronic phase, they act as powerful tags of the properties of QCD
matter, allowing for quantitative extractions of fundamental transport coefficients~\cite{Rapp:2018qla,Dong:2019byy}.

Recent measurements of the charm hadrochemistry at RHIC and the LHC, in particular the ratios $D_s/D^0$ and $\Lambda_
c/D^0$, have found intriguing patterns from $pp$ (and $p$-Pb)~\cite{Acharya:2019mgn,Acharya:2017kfy,Meninno:2019jug} to
Pb-Pb (and Au-Au)~\cite{Acharya:2018hre,Acharya:2018ckj,Vermunt:2019ecg,Sirunyan:2019fnc,Zhou:2017ikn,Adam:2019hpq}
collisions.
The $\Lambda_c/D^0$ ratio measured in $pp$ and $p$-Pb collisions at the LHC is significantly larger than that measured in
$e^+e^-$ and $ep$ collisions, and both $\Lambda_c/D^0$ and $D_s/D^0$ are seen to be further enhanced at intermediate
momenta in AA collisions. These enhancements are not easily explained by Monte-Carlo event generators based on string
fragmentation (for $pp$)~\cite{Maciula:2018iuh} and by conventional coalescence models for $AA$
collisions~\cite{Oh:2009zj,Plumari:2017ntm}, and thus provide a unique opportunity for a deeper understanding of
charm-quark hadronization~\cite{Dong:2019byy}. In the following, we report on our recent developments on charm
hadronization, mostly focusing on the transverse-momentum ($p_T$) dependent charm-hadron ratios in both $pp$ and $AA$
collisions. For $pp$ collisions, we show that the surprisingly large $\Lambda_c/D^0$ ratio can be largely explained
within an augmented statistical hadronization model (SHM) that accounts for ``missing" states to the
charm-baryon spectrum~\cite{He:2019tik} which are well motivated by both relativistic quark models (RQMs) and lattice
QCD (lQCD). For $AA$ collision, we extend our previoulsy employed momentum conserving resonance recombination model (RRM)
to the formation of charm baryons. In particular, an event-by-event implementation of the RRM allows to incorporate
space-momentum correlations (SMCs) in charm- and light-quark distributions that, together with the additional charm-baryon
states, are essential to provide a  baryon-to-meson enhancement at intermediate $p_T$~\cite{He:2019vgs}.

\section{Charm hadrochemistry in $pp$ collisions}
\begin{figure}[t]
 \begin{minipage}{.48\textwidth}
  \centering
  \includegraphics[width=\textwidth,clip]{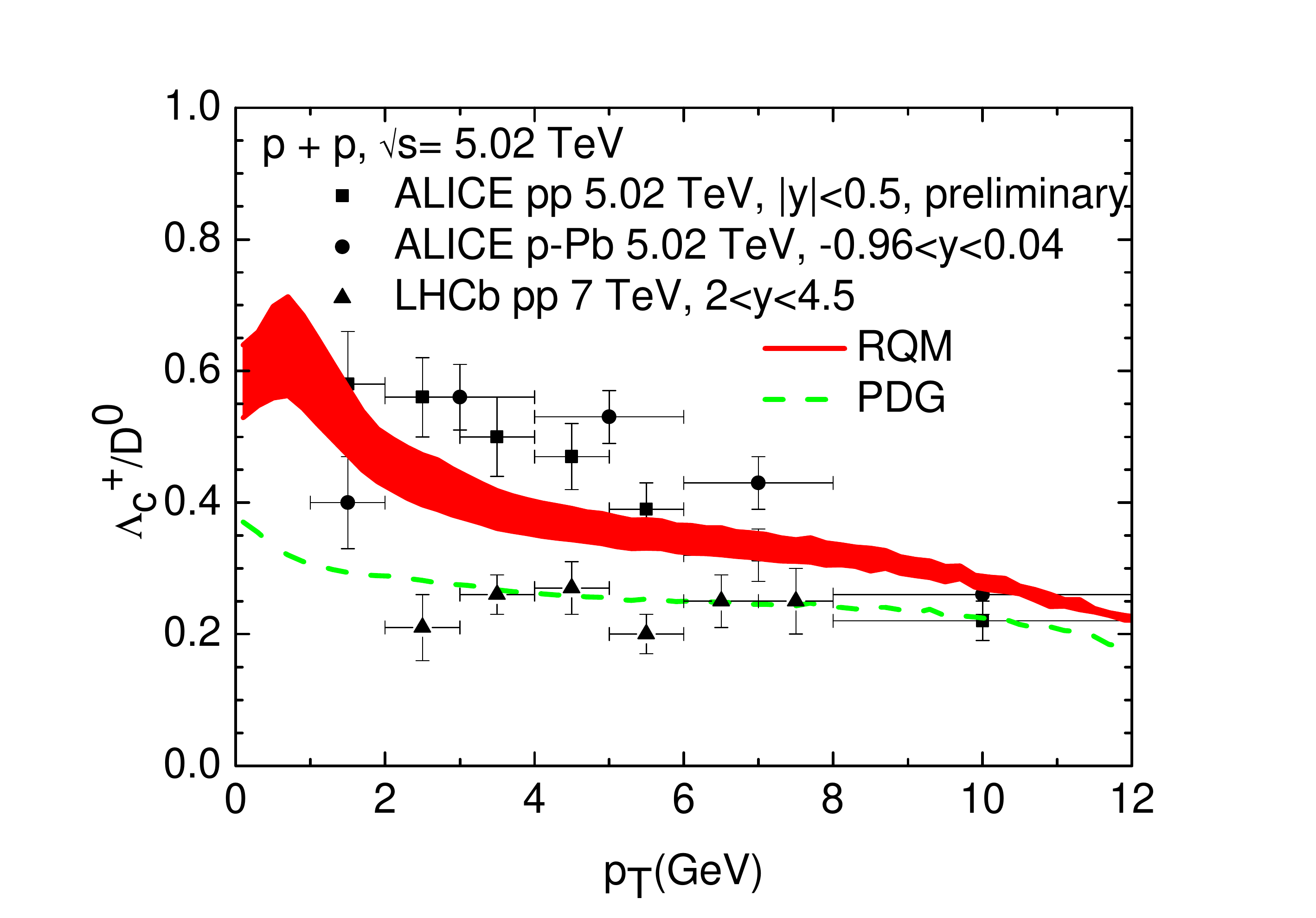}
  \end{minipage}
  \centering
  \begin{minipage}{.48\textwidth}
  \centering
  \includegraphics[width=\textwidth,clip]{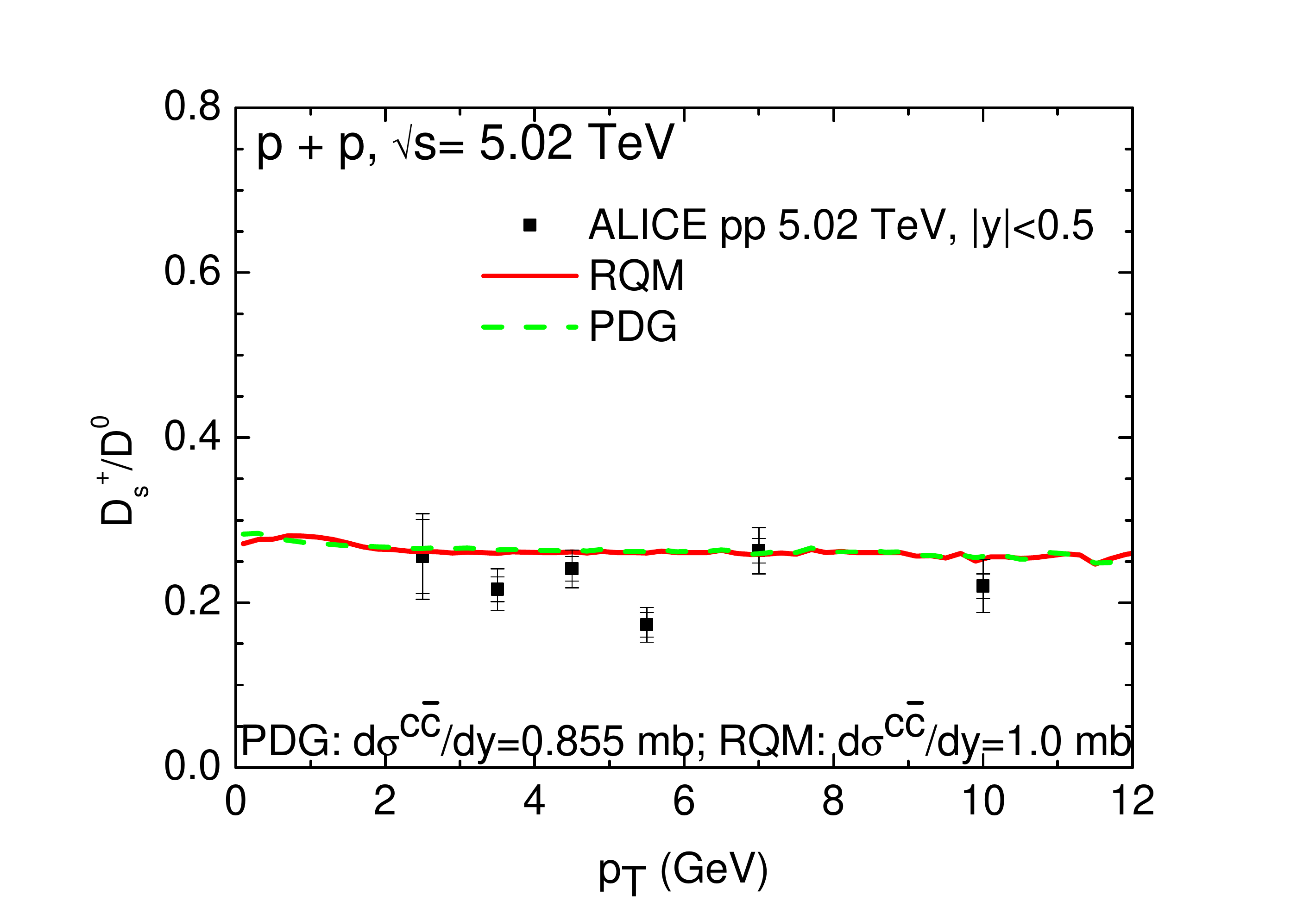}
  \end{minipage}
  \caption{The $p_T$-dependent $D_s^+/D^0$ and $\Lambda_c^+/D^0$ ratios in $\sqrt{s}=5.02$\,TeV $pp$ collisions.
The red band for $\Lambda_c^+/D^0$ in the RQM scenario represents the uncertainty in the decay branching ratios of
excited $\Lambda_c$'s and $\Sigma_c$'s into $\Lambda_c^+$ final states above the $DN$ threshold, varying between
50\% and 100\%.
  \vspace*{-0.2cm}}
  \label{fig_pp-ratios}
\end{figure}

The SHM has been successfully applied to light- and strange-hadron production in both elementary and heavy-ion collisions,
and also works for charm-meson ratios, such as $D^*/D$ or $D_s/D$ (the latter requiring a strangeness fugacity
$\gamma_s\simeq 0.6$)~\cite{Andronic:2017pug}. However, the standard SHM prediction of the cross section ratio of prompt
$\Lambda_c$ over $D^0$, $\Lambda_c/D^0\simeq 0.22$~\cite{Andronic:2007zu}, based on charm-hadron states listed by the
particle data group (PDG), is substantially below the measured value by the ALCIE collaboration in $\sqrt{s}=5$ and $7$\,TeV
$pp$ collisions~\cite{Acharya:2017kfy}. To study this puzzle, we have augmented the SHM by introducing a large set of
``missing" charm-baryon states that have not been measured (and are thus not listed by PDG) but have been predicted by
the relativistic quark model (RQM)~\cite{Ebert:2011kk}.
As a result, the $\Lambda_c^+/D^0$ ratio, at a hadronization temperature $T_H=170(160)$\, MeV reaches
$\sim 0.57(0.44)($~\cite{He:2019tik}, {\it i.e.}, a factor $\sim 2$ enhancement relative to the PDG scenario.
To calculate the $p_T$ differential cross sections, we have
performed a hadronization of the FONLL charm-quark spectrum into charmed mesons and baryons using the pertinent
fragmentation functions of the FONLL framework~\cite{Cacciari:2012ny}. The parameter $r$ in the fragmentation function for
the ground-state $D^0$ and $\Lambda_c^+$ has been tuned to fit the slope of the measured $p_T$ spectra. The spectra
of all excited states are then decayed into ground-state particles. The resulting $p_T$ differential $D_s^+/D^0$ and
$\Lambda_c^+/D^0$ ratios in $\sqrt{s}=5.02$\,TeV $pp$ collisions are shown in Fig.~\ref{fig_pp-ratios}. While both the PDG
and RQM scenarios (with a fitted total charm cross section of $d\sigma^{\rm c\bar{c}}/dy=0.855$\,mb and $1.0$\,mb, respectively)
work well for the mesonic ratio $D_s^+/D^0$, the RQM scenario with augmented charm-baryon states is favored when confronted
with the ALICE mid-rapidity data of $\Lambda_c^+/D^0$. We note that a lower ratio of about 0.35 is measured by LHCb at forward
rapidity~\cite{Aaij:2018iyy}, possibly indicating that the applicability of the SHM is compromised if the hadron multiplicity becomes
too low.

\section{Charm hadrochemistry and collectivity in $AA$ collisions}
In heavy-ion collisions, where a hot and dense QGP forms and collectively expands, the enhancement of
baryon-to-meson ratios in the light and strange sector at intermediate momenta, $p_T\simeq2-6$\,GeV,
has been attributed to quark coalescence processes. A similar enhancement has now been observed in the charm sector
by the ALICE and STAR collaborations~\cite{Vermunt:2019ecg,Adam:2019hpq}. The $p_T$-dependent modification of
the charm hadrochemistry, together with a simultaneous description of individual charm-hadron observables ($R_{\rm AA}$
and $v_2$), turns out to be a non-trivial task for theoretical modelling of charm transport and hadronization in the fireball.

In our recent work we have developed a 3-body resonance recombination model (RRM), taking advantage of the diquark
correlation in the charm-baryon sector~\cite{He:2019vgs}. In addition, within the RRM,
we have devised a method that incorporates space-momentum correlations (SMCs) between the phase space distributions
of the charm quarks (as obtained through Langevin diffusion simulations in the QGP) and light quarks (as following from
the same QGP background medium as used for the diffusion simulations). The SMCs cause, {\it e.g.}, fast moving quarks
to hadronize preferentially in the outer regions of the fireball, where the hydrodynamic flow velocity tends to be larger.
Importantly, our event-by-event implementation of the
RRM not only obeys exact charm-number conservation, but also satisfies both kinetic and chemical equilibrium limits when using thermal quark
distributions as input, resolving a long-standing problem of the fragility of coalescence model for the $v_2$ of the
formed hadrons in the presence of SMCs~\cite{Ravagli:2008rt}. These features are pivotal for controlled predictions
of the $p_T$-dependence of the charm hadrochemistry, in particular pertinent ratios.

\begin{figure}[!t]
  \centering
  \begin{minipage}{.48\textwidth}
  \includegraphics[width=\textwidth,clip]{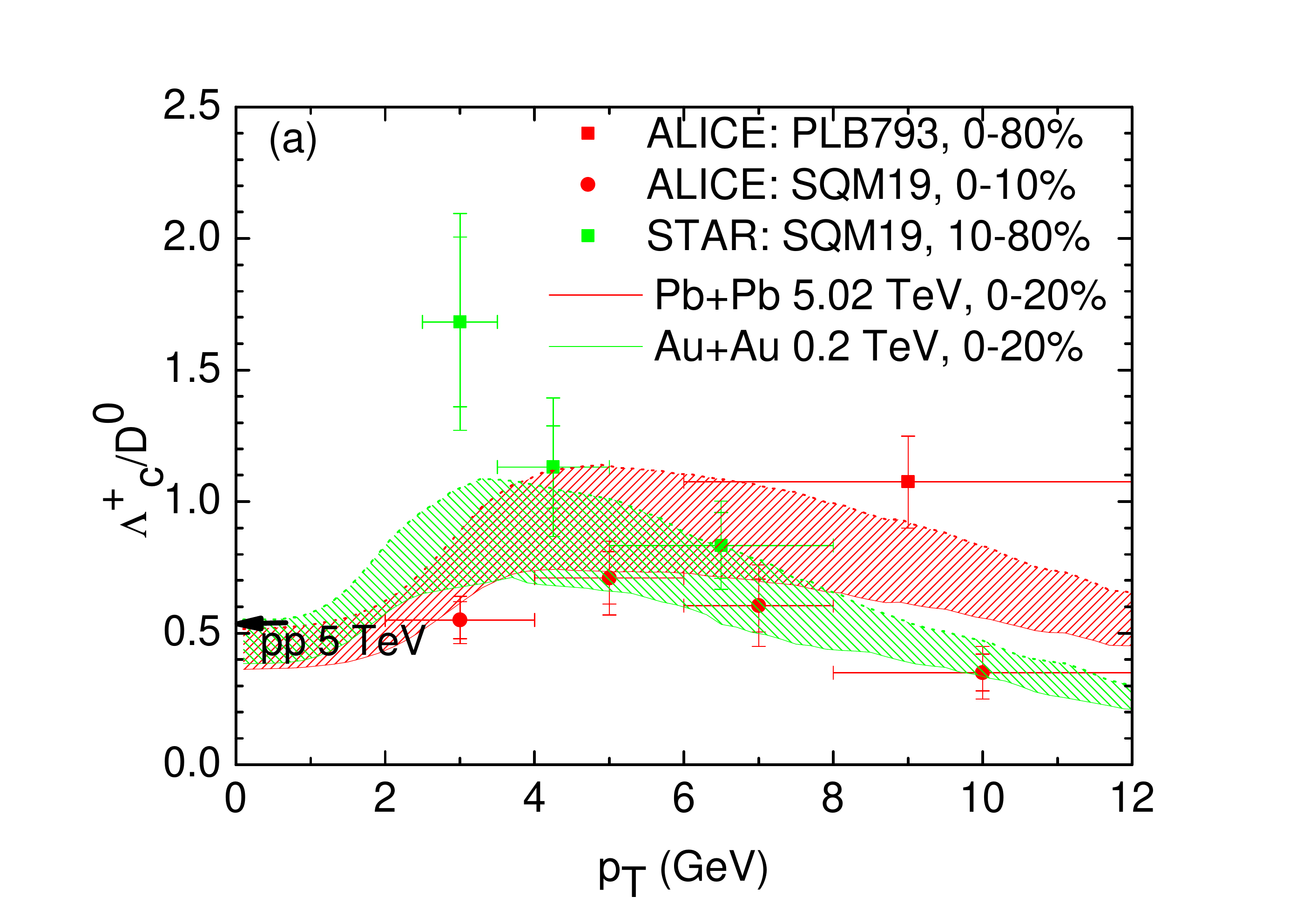}
  \end{minipage}
  \centering
  \begin{minipage}{.48\textwidth}
  \centering
  \includegraphics[width=\textwidth,clip]{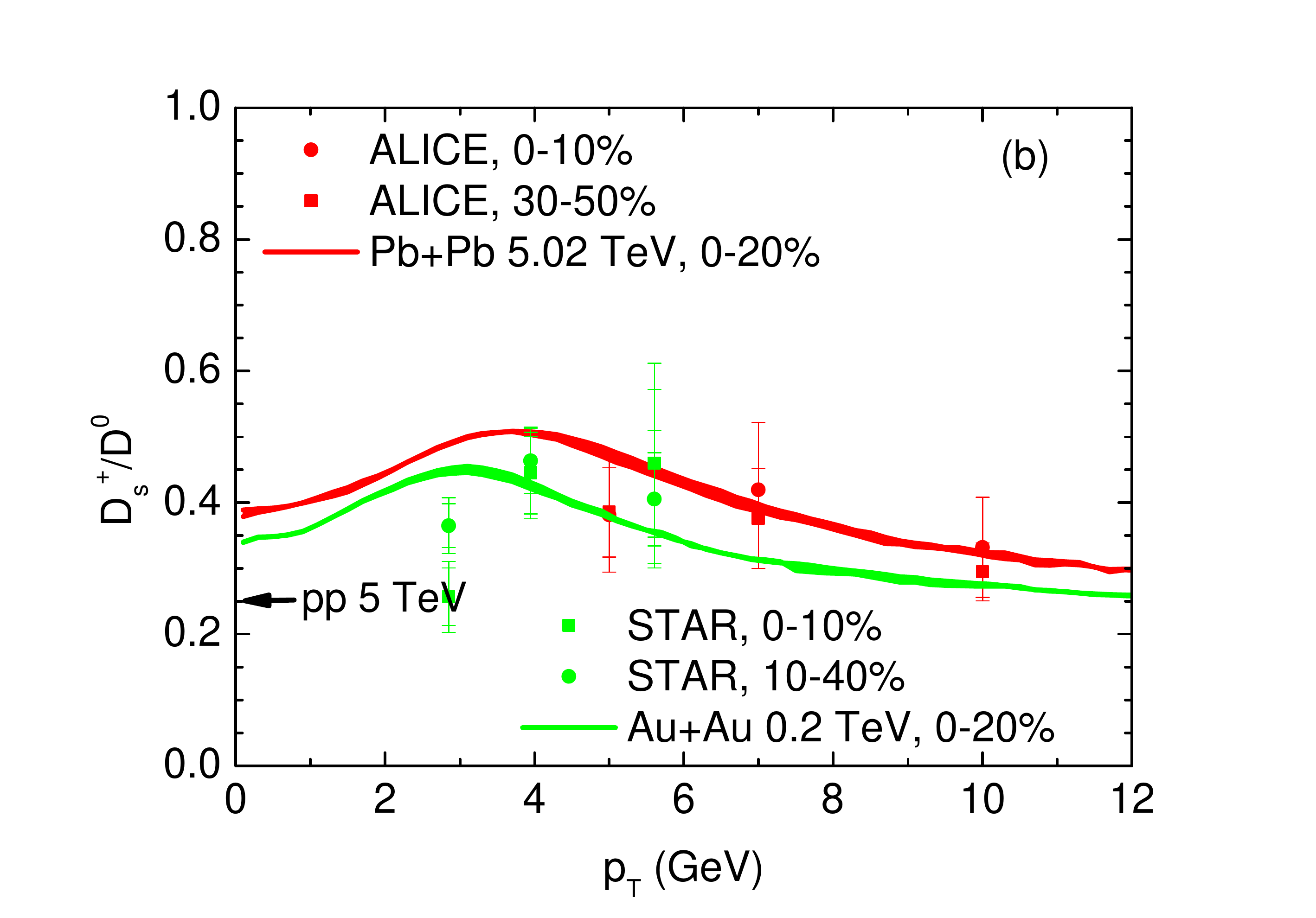}
  \end{minipage}
\caption{(a) $\Lambda_c^+/D^0$, and (b) $D_s^+/D^0$ ratio, compared to
LHC~\cite{Acharya:2018hre,Acharya:2018ckj} and RHIC~\cite{Zhou:2017ikn,Adam:2019hpq} data.
The uncertainty bands in the $\Lambda_c^+/D^0$ ratios are due to a 50-100\% BR for $\Lambda_c$ feeddown
from excited states above the $DN$ threshold~\cite{He:2019tik}, and due to hadronic diffusion in the
$D_s^+/D^0$ ratio.  The horizontal arrows indicate the $pp$ data.}
\label{fig_final-total_LcoverD0-DsoverD0}
\end{figure}

\begin{figure}[t]
\centering
  \begin{minipage}{.48\textwidth}
  \centering
  \includegraphics[width=\textwidth,clip]{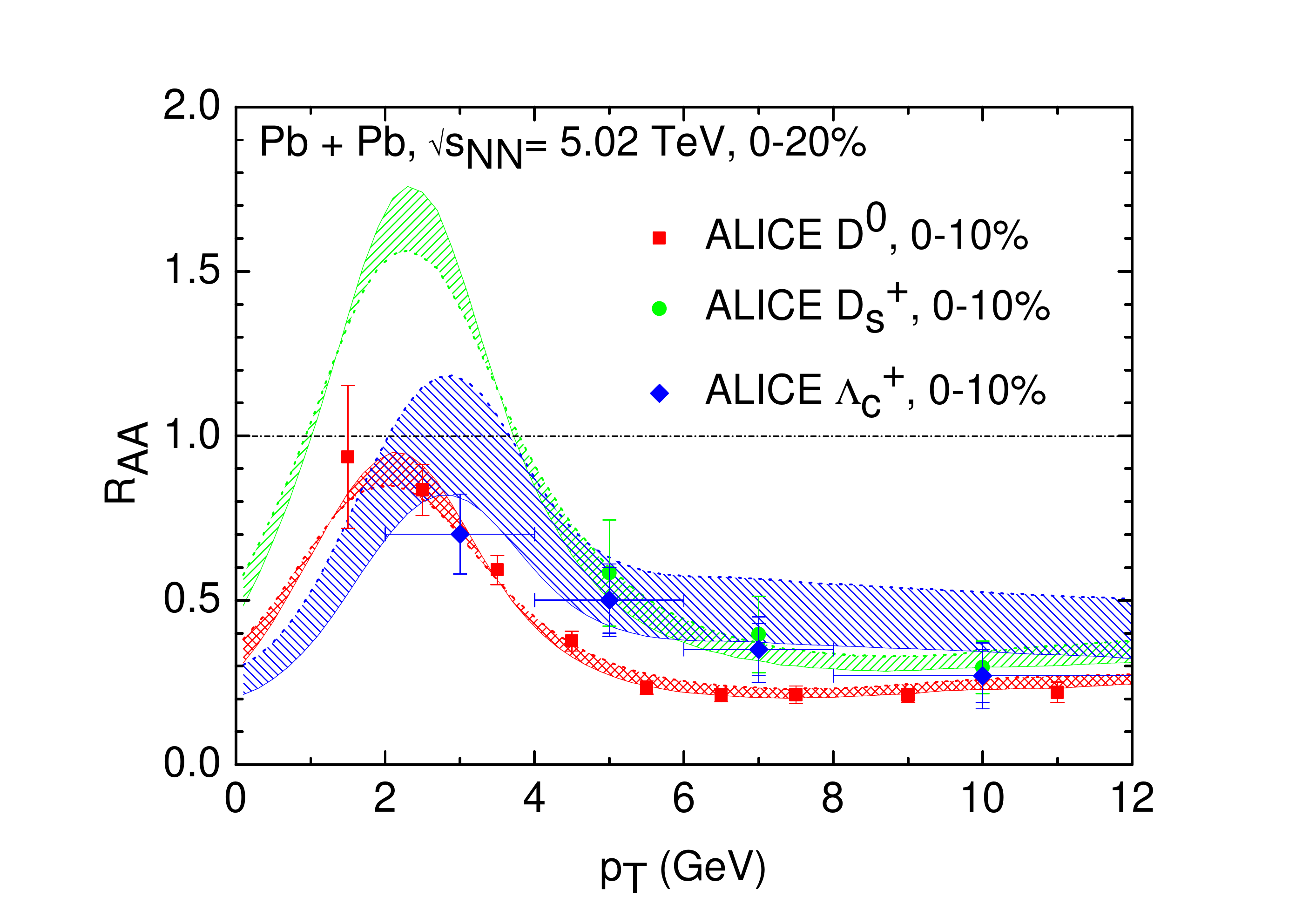}
  \end{minipage}
\begin{minipage}{.48\textwidth}
  \centering
  \includegraphics[width=\textwidth,clip]{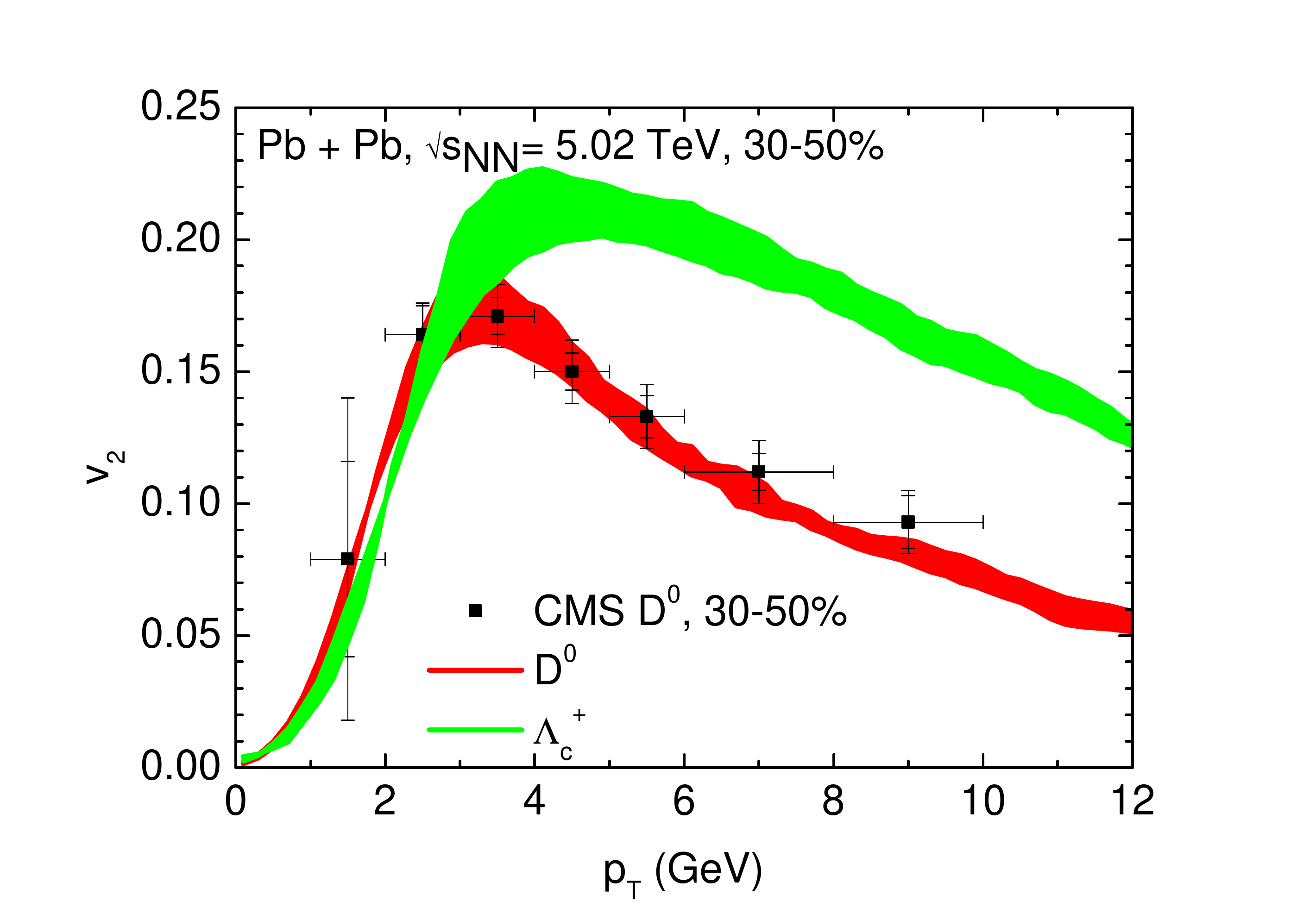}
  \end{minipage}
\caption{$R_{\rm AA}$ (left panels) and $v_2$ (right panels) of $D^0$, $D_s^+$ and $\Lambda_c^+$ in
Pb-Pb(5.02\,TeV), compared to data~\cite{Acharya:2018hre,Acharya:2018ckj,Vermunt:2019ecg}.
The uncertainty bands for the $\Lambda_c^+$ $R_{\rm AA}$ are due to a 50-100\% BR for feeddown from excited states
above the $DN$ threshold~\cite{He:2019tik}, and for the other observables due to the effects of hadronic diffusion.}
\label{fig_PbPbD0DsLc-RAA-v2}
\end{figure}

Our resulting predictions for the $\Lambda_c^+/D^0$ and $D_s^+/D^0$ ratios, including the augmented charm-baryon spectrum
as done for $pp$ baseline calculation,  are summarized in Fig.~\ref{fig_final-total_LcoverD0-DsoverD0}. The low-$p_T$
value of $\sim 0.5$ of the $\Lambda_c^+/D^0$ is essentially the same as that for $pp$, governed by the correct RRM relative
chemical equilibrium limit between baryons and mesons. The ensuing enhancement at intermediate $p_T$ is due to a stronger flow effect especially on the more massive charm baryons (feeding down to the $\Lambda_c$), as well as due to the SMCs captured by the improved RRM which render the higher (lower) $p_T$ charm/light quarks more populated in the outer (central) region of the
fireball at the time of hadronization. The overall enhancement of the $D_s^+/D^0$ relative to the $pp$ case is a result of charm
recombination in a strangeness-equilibrated environment (where the fugacity $\gamma_s$$\simeq$1)~\cite{He:2012df}.
We also note that the SMCs extend the reach of recombination
out to significantly larger $p_T$ which much improves the description of the $D$-meson $v_2$ at $p_T \gtrsim 4$\,GeV, while
the charm-hadron $R_{\rm AA}$ data are also well reproduced, cf.~Fig.~\ref{fig_PbPbD0DsLc-RAA-v2}.

\section{Conclusions}
We have investigated the charm hadrochemistry in high-energy $pp$ and heavy-ion collisions.
The role of ``missing" charm-baryons and a controlled implementation of space-momentum correlations in charm-quark recombination
processes (a long-standing problem in the field) have been highlighted, in particular the importance of recovering both kinetic
and chemical equilibrium limits in a flowing medium. The SMCs are found to significantly extend the $p_T$ reach of
recombination processes. The resulting $p_T$-dependent charm hadrochemistry and $v_2$ observables computed within our
Langevin-RRM approach show a promising degree of agreement with existing data. \\

{\it Acknowledgments.--}
This work was supported by NSFC grant 11675079 and the  U.S.~NSF under grant nos. PHY-1614484 and PHY-1913286.





\bibliographystyle{elsarticle-num}

\begin{thebibliography}{00}

\bibitem{Rapp:2018qla}
  R.~Rapp {\it et al.},
  Nucl.\ Phys.\ A {\bf 979} (2018) 21.

\bibitem{Dong:2019byy}
  X.~Dong, Y.~J.~Lee and R.~Rapp,
  Ann.\ Rev.\ Nucl.\ Part.\ Sci.\  {\bf 69} (2019) 417.


\bibitem{Acharya:2019mgn}
  S.~Acharya {\it et al.} [ALICE Collaboration],
  Eur.\ Phys.\ J.\ C {\bf 79} (2019)   388.

\bibitem{Acharya:2017kfy}
  S.~Acharya {\it et al.} [ALICE Collaboration],
  JHEP {\bf 1804} (2018) 108.

\bibitem{Meninno:2019jug}
  E.~Meninno [ALICE Collaboration],
  PoS HardProbes {\bf 2018} (2019) 137.

\bibitem{Acharya:2018hre}
  S.~Acharya {\it et al.} [ALICE Collaboration],
  JHEP {\bf 1810} (2018) 174.

\bibitem{Acharya:2018ckj}
  S.~Acharya {\it et al.} [ALICE Collaboration],
  Phys.\ Lett.\ B {\bf 793} (2019) 212.

\bibitem{Vermunt:2019ecg}
  L.~Vermunt [ALICE Collaboration],
  arXiv:1910.11738 [nucl-ex].

\bibitem{Sirunyan:2019fnc}
  A.~M.~Sirunyan {\it et al.} [CMS Collaboration],
  arXiv:1906.03322 [hep-ex].

\bibitem{Zhou:2017ikn}
  L.~Zhou [STAR Collaboration],
  Nucl.\ Phys.\ A {\bf 967} (2017) 620.


\bibitem{Adam:2019hpq}
  J.~Adam {\it et al.} [STAR Collaboration],
  arXiv:1910.14628 [nucl-ex].

\bibitem{Maciula:2018iuh}
  R.~Maciula and A.~Szczurek,
  Phys.\ Rev.\ D {\bf 98} (2018)  014016.

\bibitem{Oh:2009zj}
  Y.~Oh, C.~M.~Ko, S.~H.~Lee and S.~Yasui,
  Phys.\ Rev.\ C {\bf 79} (2009) 044905.

\bibitem{Plumari:2017ntm}
  S.~Plumari, V.~Minissale, S.~K.~Das, G.~Coci and V.~Greco,
  Eur.\ Phys.\ J.\ C {\bf 78} (2018)  348.

\bibitem{He:2019tik}
  M.~He and R.~Rapp,
  Phys.\ Lett.\ B {\bf 795} (2019) 117.


\bibitem{He:2019vgs}
  M.~He and R.~Rapp, Phys. Rev. Lett. (2020) in press;
  arXiv:1905.09216 [nucl-th].

\bibitem{Andronic:2017pug}
  A.~Andronic, P.~Braun-Munzinger, K.~Redlich and J.~Stachel,
  Nature {\bf 561} (2018)  321.

\bibitem{Andronic:2007zu}
  A.~Andronic, P.~Braun-Munzinger, K.~Redlich and J.~Stachel,
  Phys.\ Lett.\ B {\bf 659} (2008) 149.

\bibitem{Ebert:2011kk}
  D.~Ebert, R.~N.~Faustov and V.~O.~Galkin,
  Phys.\ Rev.\ D {\bf 84} (2011) 014025.


\bibitem{Cacciari:2012ny}
  M.~Cacciari, S.~Frixione, N.~Houdeau, M.~L.~Mangano, P.~Nason and G.~Ridolfi,
  JHEP {\bf 1210} (2012) 137.  

\bibitem{Aaij:2018iyy}
  R.~Aaij {\it et al.} [LHCb Collaboration],
  JHEP {\bf 1902} (2019) 102.

\bibitem{Ravagli:2008rt}
  L.~Ravagli, H.~van Hees and R.~Rapp,
  Phys.\ Rev.\ C {\bf 79} (2009) 064902

\bibitem{He:2012df}
  M.~He, R.~J.~Fries and R.~Rapp,
  Phys.\ Rev.\ Lett.\  {\bf 110} (2013)  112301






\end{thebibliography}

\end{document}